\documentstyle[preprint,eqsecnum,prd,aps]{revtex}
\begin{document} 
\preprint{
\vbox{
\halign{&##\hfil\cr
	& ANL-HEP-PR-96-78 \cr
	& September 16, 1996 \cr}}}
\title{Constraints on $\Delta G$ from Prompt Photon plus Jet Production
at HERA-${\rm \vec{N}}$ }

\author{L. E. Gordon}
\address{High Energy Physics Division, Argonne National Laboratory,
	Argonne, IL 60439}
\maketitle
\begin{abstract} 
The utility of prompt photon plus associated jet production for constraining 
the size and shape of the polarized gluon density of the proton $\Delta
G$ is examined at $\sqrt{S}=40$ GeV, appropriate for the proposed 
HERA-$\vec{{\rm N}}$ polarized $\vec{p}\vec{p}$ collider experiment. The 
calculation  is performed at next-to-leading order
(O($\alpha\alpha_s^2$)) in QCD. The 
reliability of the predictions are examined in some detail.
\vspace{0.2in}
\end{abstract}
\narrowtext

\section{ }
Polarized deep inelastic scattering (DIS) experiments have proved 
invaluable for providing information on the spin-dependent parton densities 
of the nucleon \cite{emc,smc,slac} through measurement of the polarized 
structure functions such as $g^p_1(x,Q^2)$. Data from such experiments have
traditionally been used in phenomenological analyses which attempt to
constrain the polarized parton densities and evolve them in $Q^2$
through the use of the spin-dependent Altarelli-Parisi evolution equations. 
Up until fairly recently, all such analyses had to be performed in leading order
(LO) QCD since the spin-dependent two loop splitting functions were
unknown. Since the two-loop splitting functions have been calculated
\cite{mertig,vogelsang}, recent analyses have been performed in
next-to-leading order (NLO) QCD \cite{grsv,gs,forte} and parametrizations of
the polarized parton densities have been provided which can be used  
in calculations of other polarized processes. 

In these analyses extensive use is made of theoretical constraints such as 
the Bjorken sum rule in order to extract the parton densities from the 
polarized structure function measurements. Despite this the data does not 
provide tight constraints on sizes, and even more so the shapes of either the 
polarized quark 
$(\Delta q)$ or gluon $(\Delta G)$ densities. This is true in particular for 
$\Delta G$. Each group provides more than one different parametrization of 
the parton densities, each set fitting the structure function data, but they 
can have very different individual parton densities reflecting the limited 
constraints on these quantities provided by DIS experiments. It is thus very 
important, if our understanding  of the spin structure of the nucleon is to be 
improved, that accurate  measurements of these distributions be obtained from 
other processes.      

Since the cross section for prompt photon production is dominated by the 
subprocess $qg\rightarrow\gamma q$ in hadronic collisions already in leading 
order, it has proved very useful for providing information on the 
unpolarized  gluon densities, $g(x,Q^2)$, of hadrons at fixed target 
facilities. It has therefore been suggested that it may prove equally useful 
in pinning down the polarized gluon densities \cite{bergerqiu}. In this 
context it has been examined quite extensively. The NLO corrections were 
calculated in \cite{contogouris} and \cite{gorvogel}, where numerical estimates
were also presented and it was established 
that the LO results were perturbatively stable. In both those analyses
only LO polarized parton densities were available and the
fragmentation contributions were not included. More recently, prompt
photon production with polarized beam and target was examined in
ref.\cite{gorvogel1,gorvogel2} using NLO polarized parton densities and 
including the fragmentation contributions. 

Although single inclusive prompt photon production will definitely be very 
important for constraining the size of $\Delta G$, information on the detailed 
$x$-shape of the distribution may not 
be as easily extracted. This is because the calculation of the inclusive
cross section involves one convolution over the momentum fractions, $x$, of 
the initial partons. The practical effect of this is that a measurement of
the kinematic variables of the photon is not sufficient to determine
the value of $x$. If, on the other hand, one or more of the jets
produced in the reaction is also tagged, no convolution is involved in
the calculation and the cross section is directly proportional to the
parton densities. Thus the double longitudinal asymmetry $A_{LL}$ is directly 
proportional to the ratio $\Delta G/g$ \cite{nowak} in kinematic regions
where other subprocesses such as $q\bar{q}$ scattering can be neglected.

Schematically the cross section
is given in terms of the polarized and unpolarized hard subprocess cross
sections $(\Delta)\hat{\sigma}_{ij}$ and parton densities $(\Delta)
f^a_i(x_a,M^2)$ (the presence of the $\Delta$ before a quantity means the
polarized version of the quantity) by
\begin{equation}
(\Delta)\sigma=\sum_{i,j}(\Delta)f^a_i(x_a,M^2)*(\Delta)f^b_j(x_b,M^2)*
\left[ (\Delta)\hat{\sigma}^{dir}_{ij}+\sum_{c}
(\Delta)\hat{\sigma}^{frag}_{ij\rightarrow
c}*D^{\gamma}_c(z,M_F^2)\right].
\end{equation}
The $*$ indicates either a product or convolution and explicit
dependence of the hard subprocess cross sections on the kinematic 
variables of the observed products and the renormalization and
factorization scales have been omitted. The sum over $i,j$ extends over
all initial partons and $D^\gamma_c(z,M_F^2)$ is the
unpolarized fragmentation function for a parton, $c$, into a photon at
scale $M_F^2$. Thus, just as in the single photon case, there are two classes 
of contributions to the cross section labelled the $direct$ and
$fragmentation$ contributions. 

The polarized and unpolarized parton densities are defined by
\begin{equation}
\Delta f_i(x,M^2)=f_i^+(x,M^2)-f^-_i(x,M^2)
\end{equation}
and
\begin{equation}
f_i(x,M^2)=f_i^+(x,M^2)+f^-_i(x,M^2),
\end{equation}
where the $f^+_i$ and $f^-_i$ represent the distribution of partons of
type $i$ with positive and negative helicities respectively, with respect to 
that of the parent hadron. The hard subprocess cross sections are defined by 
\begin{equation}
\Delta\hat{\sigma}_{ij}=\frac{1}{2}\left( \sigma^{++}-\sigma^{+-}\right)
\end{equation}
and
\begin{equation}
\hat{\sigma}_{ij}=\frac{1}{2}\left( \sigma^{++}+\sigma^{+-}\right).
\end{equation}
One of the main quantities studied in polarized experiments is the
longitudinal asymmetry $A_{LL}$. This quantity gives a measure of the 
sensitivity of the process to polarization effects. The double spin
asymmetry studied in this paper is defined by
\begin{equation}
A_{LL}=\frac{\Delta\sigma}{\sigma},
\end{equation}
the ratio of the cross section for longitudinal polarization of the
incoming hadrons to the corresponding unpolarized cross section.

In LO the direct processes contributing to the cross section are
$qg\rightarrow \gamma q$ and $q\bar{q}\rightarrow \gamma g$. In a LO 
jet calculation, the simple approximation `parton = jet' is used
irrespective of the jet definition used in the experiments, hence
either the final state gluon or quark will form the jet. It is only in
NLO that more than one parton may make up a jet, and one finds a
dependence of the cross section on the jet definition and parameters.
In addition there are the fragmentation processes:
\begin{eqnarray}
qg &\rightarrow& q g \nonumber \\
qq &\rightarrow& q q \nonumber \\
qq' &\rightarrow& q q' \nonumber \\
q\bar{q} &\rightarrow& q \bar{q} \nonumber \\
qg &\rightarrow& q g \nonumber \\
q\bar{q} &\rightarrow& g g \nonumber \\
gg &\rightarrow& g g \nonumber \\
gg &\rightarrow& q \bar{q} 
\end{eqnarray}
where one of the final state partons fragments to produce the photon,
i.e., $q (g)\rightarrow \gamma + X$.

The cross section of interest here is the triple differential cross
section 
\begin{displaymath}
\frac{d^3\Delta\sigma^{\gamma J}}{dp_T^\gamma d\eta^\gamma d\eta^J},
\end{displaymath}
where $\eta^\gamma$ and $\eta^J$ are the pseudorapidities of the photon
and jet respectively and  $p_T^\gamma$ is the transverse momentum of the
photon. The direct contribution to the cross section is given in LO by
\begin{equation}
\frac{d^3\Delta\sigma^{\gamma J}}{dp_T^\gamma d\eta^\gamma d\eta^J }=
2\pi p_T^\gamma\sum_{i,j}x_a\Delta f^a_i(x_i,M^2) x_b \Delta
f^b_j(x_j,M^2)\frac{d\Delta\hat{\sigma}_{ij}}{d\hat{t}}.
\end{equation}
The corresponding fragmentation
contributions will involve a convolution over the fragmentation variable
$z$. $\hat{s}=x_a x_b S$ where $\sqrt{S}$ is the cms energy in the
hadron-hadron system and as usual, $\hat{t}=(p_1-p_{\gamma})^2$ where
$p_1$ and $p_{\gamma}$ are the momenta of one of the initial state
partons and the observed final state trigger photon respectively.
The Bjorken variables, $x_{a,b}$, are given in terms of the kinematic
variables of the photon and jet by
\begin{eqnarray}
x_a&=&\frac{p^\gamma_T}{\sqrt{S}}\left(
e^{\eta^\gamma}+e^{\eta^J}\right)\nonumber \\
x_b&=&\frac{p^\gamma_T}{\sqrt{S}}\left(e^{-\eta^\gamma}+e^{-\eta^J}\right).
\end{eqnarray}
In leading order $p^\gamma_T=p_T^J$ and thus a measurement of 
$p_T^\gamma$, $\eta^\gamma$ and $\eta^J$ means that $x_{a,b}$ can be
determined. In other words, a measurement of 
$d^3\Delta\sigma^{\gamma J}/dp_T^\gamma d\eta^\gamma d\eta^J $ corresponds 
to a measurement of $d^2\Delta\sigma^{\gamma J}
/d x_a d x_b$. In NLO due to the presence of a third parton in the process
this simple formula is no longer applicable. In that case, in order to
determine $x_{a,b}$ unambiguously, one would need to measure the
kinematic variables of all three final state partons. The values of
$x_{a,b}$ are only approximately determined by the transverse momenta
and pseudorapidities of the photon and jet in this case, due to the need
to implement a jet definition. If photon isolation requirements are also
implemented, then the determination of $x_{a,b}$ from the photon and jet
kinematic variables would be even less precise. Such a
measurement would still nevertheless yield some very useful information
on the $x$-dependence of the polarized parton densities as will be
shown.

In  NLO, $O(\alpha\alpha_s^2)$, there are virtual corrections 
to the LO non-fragmentation processes, as well as the further 
three-body processes:
\begin{mathletters}\label{eq:1}
\begin{eqnarray}
q &+g  \rightarrow q + g + \gamma\label{eq:11}\\
g &+ g \rightarrow q +\bar{q} + \gamma\label{eq:12}\\
q &+ \bar{q} \rightarrow g +g + \gamma\label{eq:13}\\
q &+ q \rightarrow  q + q + \gamma\label{eq:14}\\
\bar{q} &+ q \rightarrow \bar{q} + q + \gamma\label{eq:15}\\
q &+ \bar{q} \rightarrow q' + \bar{q}' + \gamma\label{eq:16}\\
q &+ q' \rightarrow q' + q + \gamma\label{eq:17}
\end{eqnarray}
\end{mathletters} 
The virtual corrections as well as the three-body matrix elements were
calculated in ref.\cite{gorvogel} and I use these matrix elements in
this calculation.

In principle the the fragmentation processes of eq.(1.7) should now be
calculated to $O(\alpha_s^3)$ and convoluted with the NLO photon
fragmentation functions whose leading behavior is $O(\alpha/\alpha_s)$,
but the hard subprocess matrix elements are not yet available in the
polarized case. Hence unless otherwise stated, in both the polarized and 
unpolarized cases, I include the leading order contributions to these 
processes only. Numerically the fragmentation processes are not significant 
except at extremely low $p_T$ due to the low cms energy of the fixed
target experiment as I shall show later, but for a theoretically consistent
calculation they should nevertheless be included as they help to reduce
uncertainties from scale dependence.

For this calculation I use the Monte Carlo method first used in
\cite{ohnemus} for the unpolarized case of the photon plus jet calculation 
performed here, as well as for calculations of some other important
processes, to deal 
with the phase space integrals. In this letter I shall not
give any details of the calculation except to say that the calculation
was performed in the $\overline{MS}$ scheme using the t'Hooft-Veltman
(HVBM) scheme \cite{hvbm} to treat $\gamma_5$. I refer to ref.\cite{gordon} 
for the 
details of the calculation. At NLO, since there are two-to-three scattering 
processes included in the calculation it is possible that two partons may be too
close together to be resolved as two separate jets. In this case a jet 
definition is required. In this calculation I use the cone definition proposed 
at Snowmass \cite{snowmass}, which defines a jet as hadronic energy deposited 
in a cone of radius $R_J=\sqrt{(\Delta\eta)^2+(\Delta\phi)^2}$. At the partonic
level, if one parton forms the jet then the kinematic variable of the
parton is set equal to that of the jet, $p_J=p_i$, $\eta_J=\eta_i$ and
$\phi_J=\phi_i$, and the jet is centered on the parton. If another
parton falls inside the jet cone then the jet variables are the weighted
averages of those of the two partons:
\begin{eqnarray}
p_J&=&p_1+p_2 \nonumber \\
\eta_J&=&\frac{1}{p_J}\left(p_1\eta_1+p_2\eta_2\right)\nonumber \\
\phi_J&=&\frac{1}{p_J}\left(p_1\phi_1+p_2\phi_2\right).
\end{eqnarray}

All results are displayed for $\vec{p}\vec{p}$ collisions at the center-of-mass
energy $\sqrt{s}=40$ GeV appropriate for the proposed HERA-$\vec{\rm N}$ 
experiment at DESY \cite{nowak}. For the unpolarized cross section the
CTEQ3M parton densities are used throughout, and the value of
$\Lambda_{\overline{MS}}$ corresponding to this distribution is also used.
Use of other unpolarized parton densities at the $x$-values probed
here do not yield significantly different results.
For the polarized case the GRSV \cite{grsv} and GS \cite{gs}
distributions are used with the corresponding values for 
$\Lambda_{\overline{MS}}$. The authors of ref.\cite{grsv} and \cite{gs} have 
proposed various parametrizations of the polarized parton densities
differing mainly in the choice of input for the polarized gluon density
$\Delta G$. In the case of the GRSV distributions we use the `valence'
set which corresponds to a fit of the available DIS data (referred to by
the authors as the `fitted' $\Delta G$ scenario), the large
gluon fit which assumes that $\Delta G(x,Q_0^2)=g(x,Q_0^2)$ at input (the
'$\Delta G=g$' scenario) and 
the small gluon fit which uses $\Delta G(x,Q_0^2)=0$ at the input scale
(the '$\Delta G=0$' scenario), which in this case starts at the very low
value of $Q_0^2=0.34$ GeV$^2$. The latter two distributions are intended
to represent extreme choices for $\Delta G$. 
These parametrizations give gluon densities which differ in their absolute
sizes as well as in their $x$-shape. The GS parametrizations provide
three fits to the data; GS A, GS B and GS C. It has been shown that the
GS A and GS B distributions do not differ very much from the $\Delta
G=g$ and fitted $\Delta G$ sets of GRSV respectively, whereas the the GS C 
set is widely
different from any of the others. I shall present
distributions using the three GRSV sets discussed above, along with the
GS C set for comparison. For the fragmentation functions I use
the LO asymptotic parametrization of ref.\cite{owens}. As will be shown,
the choice of fragmentation functions makes very little difference to
the predictions, since these processes account for only a small fraction
of the cross section.

The renormalization, factorization, and  fragmentation scales are set to a 
common value $\mu = p_T^{\gamma}$ unless otherwise stated.  
Dependence on $\mu$ is examined in one of the 
figures below.  Since there are two `particles' in the final state,
the jet and the photon, both of whose transverse momenta are
large, an alternative choice might be $\mu = p_T^J$ or some function
of $p_T^{\gamma}$ and $p_T^J$.  The results of the calculations
show that the magnitudes of $p_T^{\gamma}$ and $p_T^J$ tend to be 
comparable and that dependence of the asymmetries on $\mu$ is slight,
although the individual cross sections may vary significantly with $\mu$.  
Therefore, choices of $\mu$ different from $\mu = p_T^{\gamma}$ should not 
produce significantly different predictions for the asymmetries . The two loop 
expression for
$\alpha_s(Q^2)$ is used throughout, with the number of flavors fixed at
$N_f=4$, although the contribution from charm was verified to be negligible
at the energies considered. The value of the jet cone size is fixed at
$R_J=0.5$ unless otherwise stated.

Fig.1a shows the triple differential cross section as a function of 
$p^\gamma_T$ of the photon for the various parametrizations. The unpolarized 
cross section is shown for comparison. The curves were obtained 
by averaging over $\Delta \eta=1$ and $\Delta p_T^\gamma=1$ GeV and the 
restriction $p_T^J\geq p_T^\gamma$ was imposed. The upper dotted curve
is for the GS A parametrization, verifying that it is very similar in
both shape and size to the $\Delta G= g$ parametrization of GRSV. All
the remaining parametrizations give distributions which are distinctly
different in both their shapes and sizes. If one compares these
distributions with the corresponding curves for single inclusive prompt
photon production \cite{gorvogel1} one finds, apart from the expected fall in 
the absolute size of the cross section, that the corresponding distributions 
differ also in their shapes. This is most obvious for the GS C parametrization
which has the most distinctly different shape. Fig.1b also 
shows significantly larger asymmetries than
the inclusive photon case for equivalent $p_T^\gamma$'s. Most of these 
differences can be traced to the fact that for the inclusive photon
case, a given value of $p_T^\gamma$ corresponds to a much less sharply
defined value of $x$ than in the photon plus jet case. In Fig.1a,b
the range $4\leq p_T^\gamma\leq 10$ GeV corresponds to $0.2 \leq x_{a,b}
\leq 0.5$. It was also verified that for all parametrizations, the $qg$
scattering process accounts for more than $90\%$ of the cross sections.
Also plotted in fig.1b are the projected statistical errors in the
asymmetries, $\delta A_{LL}$, as estimated in ref.\cite{nowak} from the
formula
\begin{equation}  
\delta A_{LL}=0.17/\sqrt{\sigma(pb)}.
\end{equation}
In \cite{nowak} $\delta A_{LL}$ has been calculated by assuming an
integrated luminosity of $240$ pb$^{-1}$ and beam and target polarizations
of $P_B=0.6$ and $P_T=0.8$, and including a trigger and reconstruction
efficiency of $50\%$ with no acceptance correction. Similar errors for
$\delta A_{LL}$ were estimated in \cite{gorvogel1} by integrating the
unpolarized cross section over bins of sizes $\Delta \eta=1$ and $\Delta
p_T=1$ GeV. We use the same proceedure here. The results indicate that
for $p_T\leq 7-8$ GeV the asymmetries will be distinguishable.

Fig.1c and Fig.1d are similar to Figs.1a,b but in this case both the jet
and photon are restricted to the forward rapidity regions. $\eta^\gamma$
and $\eta^J$ are centered at $+1$, with $\Delta \eta=0.2$. At
$p_T^\gamma=5$ GeV this corresponds to probing $x_{a,b}$ in bins
centered at the points $x_{a,b}=0.07$ and $x_{b,a}=0.55$. The asymmetries 
are smaller than in fig.1b but
again there are clear distinctions in sizes and shapes for the different
parametrizations. 

In Fig.2 we look at distributions in $\eta^\gamma$ at $p_T^\gamma=5$ GeV 
and allow $p_T^J$ to vary between $5$ and $20$ GeV. In Fig.2a the jet is 
restricted to be in the central region, $-0.5\leq \eta^J \leq 0.5$, and in 
Fig.2c it is restricted to the forward region, $0.5\leq \eta^J\leq 1.5$. The
asymmetry plots of Fig.2b shows important differences between the various
parametrizations but that of Fig.2d is very striking. The asymmetries
are rather flat, although showing differences in shape for the various
parametrizations in the positive rapidity regions, but are very large
and increase very sharply in the negative rapidity region, almost
approaching $1$ at the edge of phase space for the $\Delta G=g$ scenario. 

An examination
of Fig.2c shows that this effect originates with the difference in
shapes between the polarized and the unpolarized rapidity distributions.
It was argued in ref.\cite{berger} that positive rapidity correlations
at collider energies
are an inherent property of the hard subprocess matrix elements. Thus if
one restricts the jet to be at positive rapidity then the rapidity distribution
of the photon would be expected to peak in the positive rapidity region. This 
was verified recently in ref.\cite{bailey} where prompt photon plus charm
quark correlations were studied at NLO. Clearly in Fig.2c this expectation
is confirmed for the unpolarized cross section, but in contrast the
polarized cross sections show {\it negative} correlations. Since the
propagators in the hard subprocess matrix elements are the same this can
only be due to the differences between the polarized and unpolarized
parton densities. This fact was verified by using artificial
polarized parton densities of exactly the same shape as the unpolarized ones 
along with the polarized matrix elements. In that case positive rapidity
correlations between the photon and the jet were obtained. 

In ref.\cite{berger} it is suggested that it is the behavior at 
small-$x_a, x_b$ 
of the product $x_a x_b f^a(x_a,Q^2) f^b(x_b,Q^2)$ along with the
structure of the hard subprocess matrix elements which generate the
positive rapidity correlations between two final state particles. It is well 
established that the small-$x$ behavior of the ratio $\Delta G/g$ is 
$\sim x$ as $x\rightarrow 0$. Hence there is an additional power of $x$ in the 
polarized parton distributions at small-$x$. This also applies in the
set for which $\Delta G=g$ at the input scale. 
At the scale we are considering ($Q^2\sim 25$ GeV$^2$) the
evolution equations determine the shape of the parton densities
regardless of the shape at the input scale. 
Therefore given the fact that
the behavior of the ratio $\Delta G/g$ at small-$x$ is generated by the 
polarized Altarelli-Parisi equations
themselves \cite{bergerqiu} and that the GRSV input scale is so low
($Q^2_0=0.34$ GeV$^2$), it is understandable that even the $\Delta G=g$
scenario will also show the same small-$x$ behavior characteristic of
the polarized parton densities.  
It is easy to show from eq.(1.9) 
\cite{berger} that when only two final state particles are present, as
is the case in LO 
\begin{equation}
x_a x_b=\frac{2 (p_T^\gamma)^2}{S}(1+\cosh(\eta^\gamma-\eta^J)).
\end{equation}
The effect of this additional factor is to suppress the polarized
rapidity distributions at the
point $\eta^\gamma=\eta^J$ and produce two symmetrical peaks on both sides of
this point. There are many other factors such as, for example, available phase
phase  which also affect the rapidity correlations and may tend to obscure the
effect of the small-$x$ behavior, but from Fig.2d the effect is a
significant rise in the predicted asymmetries at negative rapidities. 
This strong sensitivity to polarization effects in this region should
serve as a very good test of the underlying QCD mechanism as well as a check on
assumptions about the small-$x$ behavior of the polarized parton
densities.

In order to test the sensitivity of the predictions to the choice of
factorization and renormalization scales, Fig.3a shows predictions for the
asymmetry of a sample of the the $p_T^\gamma$ distributions of Fig.1a for 
three different scales. All hard scales are kept equal and varied 
simultaneously.
Although, as in the case of single prompt photon production, varying the
scales over the full range shown can change the individual cross
sections by up to $50\%$, the predicted asymmetries are relatively
stable. In Fig.3b the sensitivity of the predicted asymmetries to the
cone size of the jet $R_J$ and to the inclusion of the fragmentation
contributions is tested. The solid lines are again the predictions of
Fig.1b. Setting $D_{c/\gamma}(z,M_F^2)=0$ clearly has very little effect
on the asymmetries, meaning that the predictions are likely to be very
similar if NLO corrections to these contributions were included. This
could be expected since at this cms energy, fragmentation is less the
$10\%$ of the cross section except at the very lowest $p_T$ values. The
dotted line in Fig.3b shows the effect of setting $R_J=1$. The effect on
the cross sections is generally less than $10\%$ and clearly it has no
effect at all on the asymmetries.    

Since in this calculation a jet definition is involved in NLO, the
calculation of a so-called K-factor to estimate the size of the NLO
corrections is not meaningful, as it would depend on the value of $R_J$
chosen. Thus in Fig.3c, a comparison is made between predictions for the
cross section using the LO and NLO matrix elements but both using the NLO 
structure functions. The aim is to examine whether the NLO matrix
elements and jet definition have any effect on the shapes of the
distributions. As stated earlier, in NLO, measuring the kinematic
variables of the photon and jet does not serve to uniquely determine the
values of $x_a$ and $x_b$. Any significant differences between the
values of these variables probed in LO and NLO could possible show up as
a difference in the shapes of the distributions. Figs.3a,b clearly show
that there are no significant shape differences between the LO and NLO
distributions, and therefore one can assume that estimating the values of 
$x$ probed for a given kinematical configuration of the photon and jet by the 
LO formula may not lead to unacceptably large errors.

In conclusion, I have examined the possibility that both the size and
$x$-shape of the polarized gluon distribution of the proton, $\Delta G$, may
be measured at HERA-$\vec{\rm N}$ via a measurement of the photon plus
jet cross section. Control over the kinematic variables of both the
photon and jet allows a much better determination of the $x$-value
probed when compared to inclusive prompt photon production. A comparison
of the predictions obtained using different polarized parton densities
show that a clear distinction between both the sizes and shapes should be
possible. Assuming that the `fitted $\Delta G$' scenario is the most
plausible distribution, then a typical value for the asymmetry, $A_{LL}$
is $10\%$, but given the uncertainty in $\Delta G$ the asymmetry could
be as small as $2\%$ or as large as $40\%$. The expected small-$x$ behavior 
of the polarized distributions lead to predictions of negative correlations 
between the rapidities of the photon and jet. The effect is to produce very 
large asymmetries, even approaching the maximum value of $A_{LL}=1$ in certain
kinematic regions, which should make them more easily measurable in the 
experiments.   

\section{Acknowledgments}

I am indebted to Ed Berger for many helpful discussions and for reading the
manuscript, and to Werner Vogelsang for some helpful comments.
The work at Argonne National Laboratory was supported by the US Department of
Energy, Division of High Energy Physics, Contract number W-31-109-ENG-38.
\pagebreak


\pagebreak

\noindent
\begin{center}
{\large FIGURE CAPTIONS}
\end{center}
\newcounter{num}
\begin{list}%
{[\arabic{num}]}{\usecounter{num}
    \setlength{\rightmargin}{\leftmargin}}

\item (a) $p_T^\gamma$ distribution of the photon plus jet triple
differential  cross section $d^3\sigma^{\gamma J}/dp_T^\gamma d
\eta^\gamma d\eta^J$ for various polarized parton distributions for
rapidities of the photon and jet averaged over the region 
$-0.5\leq\eta^\gamma,\eta^J\leq 0.5$. The
upper dotted curve is for the GS A polarized distributions. (b)
Longitudinal asymmetries for the distributions of (a). (c) and (d)
similar to (a) and (b) but for both the photon and jet
rapidities averaged over the region  
$0.5\leq\eta^\gamma,\eta^J\leq 1.5$.
\item (a) Distribution in the rapidity of the photon for $p_T^\gamma=5$
GeV and $\eta^J$ averaged over the region $-0.5\leq\eta^J\leq0.5$ and 
$p_T^J\geq p_T^\gamma$. (b) Asymmetries for the curves shown in (a). (c) and 
(d), similar to (a) and (b) respectively, but for $\eta^J$ averaged over 
$0.5\leq \eta \leq 1.5$. 
\item (a) Scale dependence of the asymmetries shown in Fig.1b, for three
parametrizations of the polarized densities. (b) Comparison of the asymmetries 
of Fig.1b after setting $R_J=1.0$ (dotted lines) and on the same plot,
after neglecting the contributions from fragmentation (dashed lines). 
(c) Comparison of the triple differential cross section of Fig.1a with
those using the LO matrix elements and (d) the corresponding
asymmetries.
\end{list}
\pagebreak


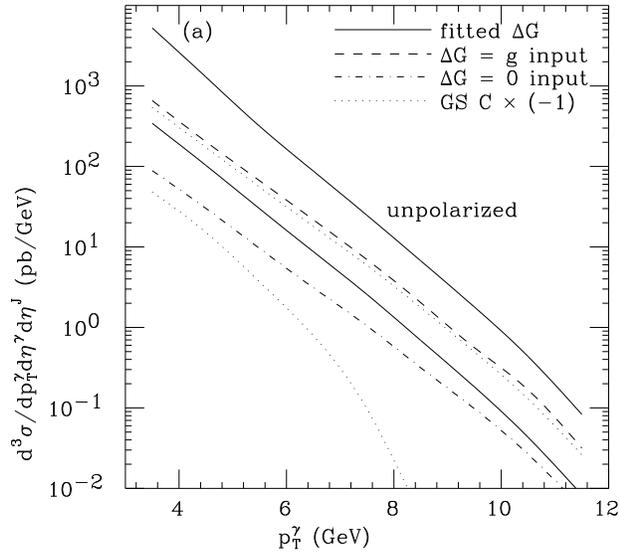
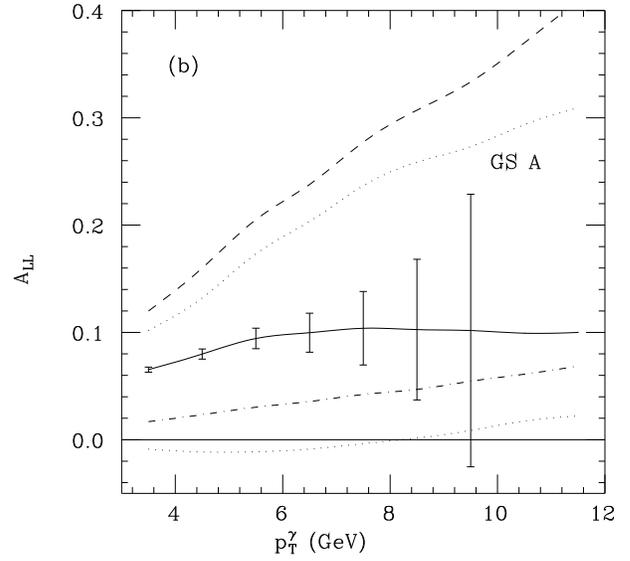
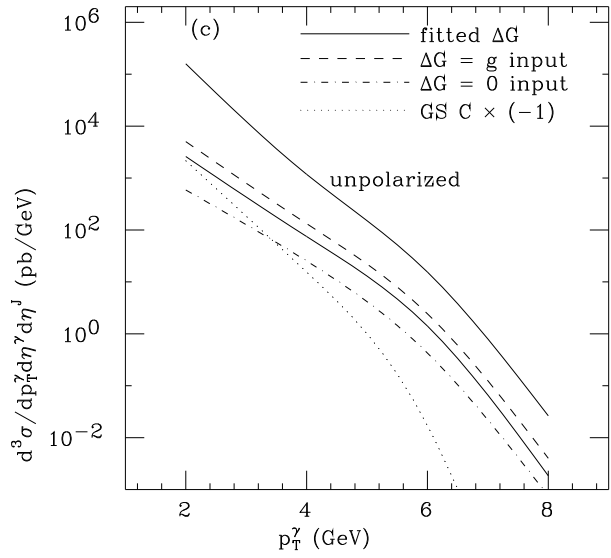
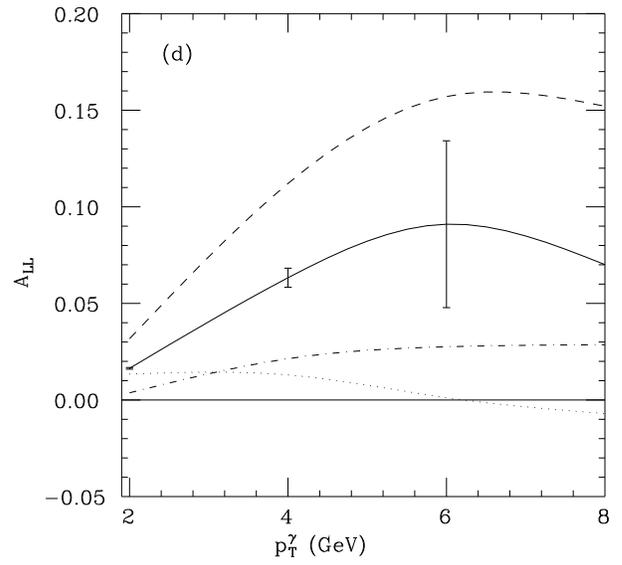
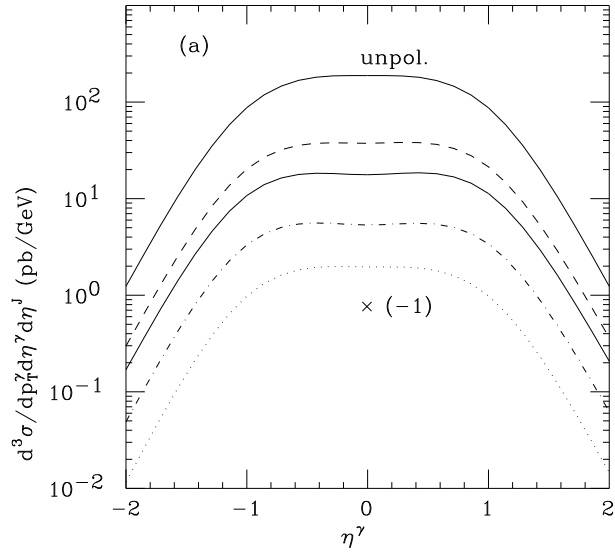
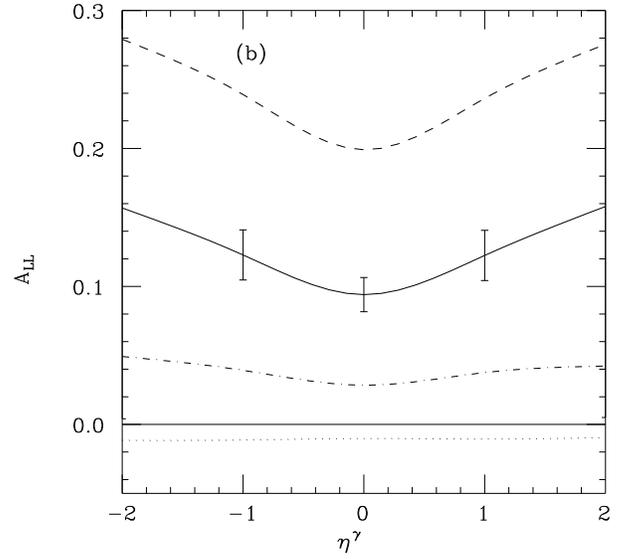
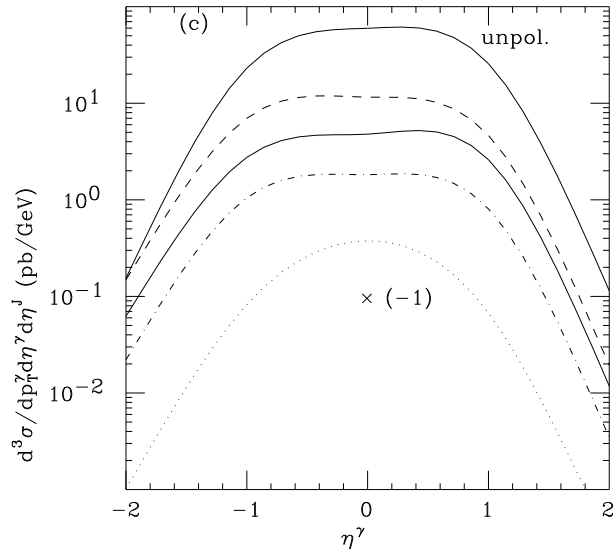
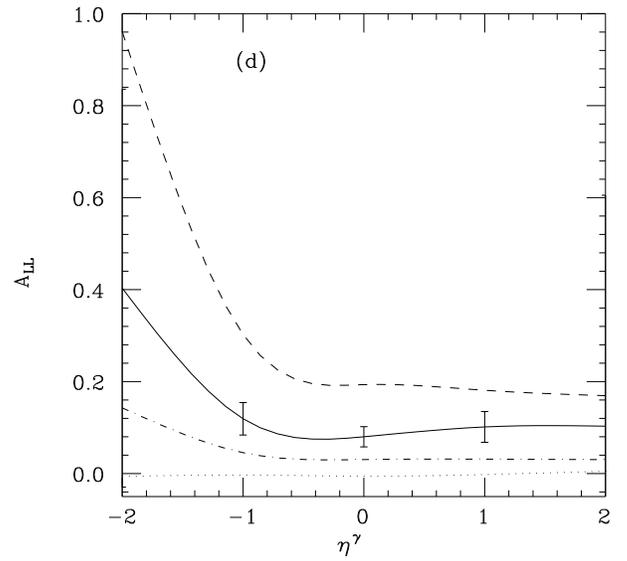
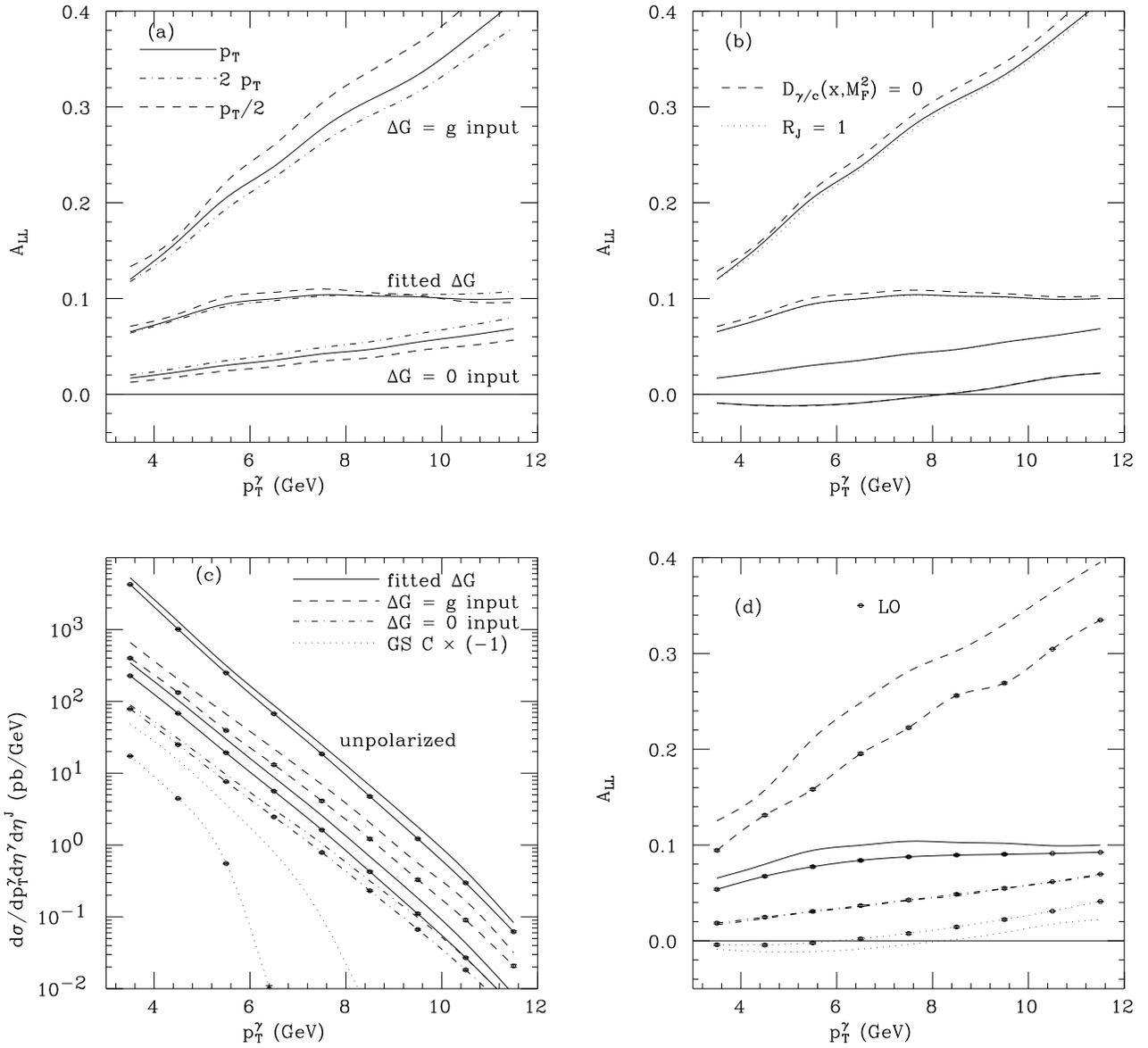
\end{document}